\documentclass[10pt, conference]{IEEEtran}
\IEEEoverridecommandlockouts
\usepackage[dvipsnames]{xcolor}
\usepackage{cite}
\usepackage{amsmath,amssymb,amsfonts}
\usepackage{makecell}
\usepackage[inkscapeformat=png]{svg}
\usepackage{graphicx}
\usepackage{textcomp}
\usepackage{amsthm,amssymb}
\usepackage{epstopdf}
\usepackage{url}
\usepackage{wrapfig}
\usepackage{array}
\newcommand{\PreserveBackslash}[1]{\let\temp=\\#1\let\\=\temp}
\newcolumntype{C}[1]{>{\PreserveBackslash\centering}p{#1}}
\newcolumntype{R}[1]{>{\PreserveBackslash\raggedleft}p{#1}}
\newcolumntype{L}[1]{>{\PreserveBackslash\raggedright}p{#1}}

\ifCLASSOPTIONcompsoc
\usepackage[caption=false,font=normalsize,labelfon
t=sf,textfont=sf]{subfig}
\else
\usepackage[caption=false,font=footnotesize]{subfig}
\fi
\usepackage[algoruled, vlined, linesnumbered]{algorithm2e}  
\usepackage{booktabs}

\SetKwInOut{Parameter}{Parameters}

\makeatletter
\def\endthebibliography{%
	\def\@noitemerr{\@latex@warning{Empty `thebibliography' environment}}%
	\endlist
}
\makeatother

\DeclareMathAlphabet{\mathcalboondox}{U}{BOONDOX-calo}{m}{n}
\SetMathAlphabet{\mathcalboondox}{bold}{U}{BOONDOX-calo}{b}{n}
\DeclareMathAlphabet{\mathbcalboondox}{U}{BOONDOX-calo}{b}{n}

\def\BibTeX{{\rm B\kern-.05em{\sc i\kern-.025em b}\kern-.08em
		T\kern-.1667em\lower.7ex\hbox{E}\kern-.125emX}}

\begin{document}
	
	\title{PDCCH Scheduling via Maximum Independent Set}

	\author{Lorenzo Maggi, Alvaro Valcarce Rial, \emph{Senior IEEE member}, Aloïs Herzog, Suresh Kalyanasundaram, Rakshak Agrawal 
		
		\thanks{Lorenzo Maggi and Alvaro Valcarce Rial are with Nokia Networks France, Bell Labs, 91377 Massy Palaiseau (France).  Alo\"is Herzog is with Nokia Networks France, Mobile Networks, 91377 Massy Palaiseau (France). Suresh Kalyanasundaram and Rakshak Agrawal are with Nokia India Private Limited, Mobile Networks, 560045 Bangalore (India). Emails: \{lorenzo.maggi, alvaro.valcarce\_rial\}@nokia-bell-labs.com, \{alois.herzog, suresh.kalyanasundaram, rakshak.agrawal\}@nokia.com}}
	
	
	
	\maketitle

	\begin{abstract}
		In 5G, the Physical Downlink Control CHannel (PDCCH) carries crucial information enabling the User Equipment (UE) to connect in UL and DL. UEs are unaware of the frequency location at which PDCCH is encoded, hence they need to perform blind decoding over a limited set of possible candidates. 
		We address the problem faced by the gNodeB of selecting PDCCH candidates for each UE to optimize data transmission. 
		We formulate it as a Maximum Weighted Independent Set (MWIS) problem, that is known to be an NP-hard problem and cannot even be approximated.
		A solution method called Weight-to-Degree Ratio (WDR) Greedy emerges as a strong contender for practical implementations due to its favorable performance-to-complexity trade-off and theoretical performance guarantees.	
	\end{abstract}
	
	
	\section{Introduction}
	
	In 5G wireless communication systems, the Physical Downlink Control CHannel (PDCCH) comprises the set of time/frequency resources carrying Downlink Control Information (DCI). 
	Each DCI, specific to either downlink (DL) or uplink (UL), enables UEs to establish communication in the upcoming slot, and conveys information on scheduling assignments, slot format indication, power control, precoding, modulation and coding schemes. 
	A UE is unaware of the specific sub-carriers where its DCI has been encoded, thus it performs blind decoding. To ease this process, the UE searches within a restricted set of candidate frequency locations, known as Search Space (SS). However, this complicates the gNodeB's task of selecting PDCCH candidates and maximizing the resource allocation fairness across UEs.
	
	To define more precisely a SS we need to introduce some terminology compliant with the 3GPP specifications \cite{TS38211}. 
	
\begin{figure}
		\centering
		\includegraphics[width=\linewidth]{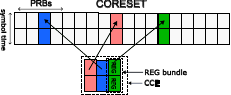}
		\caption{Example of interleaved CCE-to-REG mapping in a CORESET, where CORESET spans 2 OFDM symbol times.}
		\label{fig:coreset}
	\end{figure}

	\emph{Preliminaries.} A RE denotes one subcarrier and 1 Orthogonal Frequency Division Multiplexing (OFDM) symbol time, carrying 1 OFDM symbol. A Resource Element Group (REG) is a set of 12 REs contiguous in frequency, being equivalent to a Physical Resource Block (PRB) over one OFDM symbol time. 
	REGs can be grouped over one or two consecutive symbol times into REG \emph{bundles}. The latter notion finally allows us to introduce the smallest logical building block of a SS, called Control Channel Element (CCE) and consisting of 6 REGs. If REGs are bundled in two slots, then a CCE is a set of 3 REG bundles, each of which is interleaved in frequency. The set of PRBs and OFDM symbols on which CCEs are allocated is called COntrol REsource SET (CORESET), see Figure \ref{fig:coreset}.
	The code rate used for DCI transmission depends on the channel conditions experienced by the UE: the worse the channel conditions, the lower code rate. Thus, a variable number of CCEs, called Aggregation Level (AL), is used to encode a single DCI. Possible values of AL are $\{1,2,4,8,16\}$.

	\emph{Search Space (SS).} A DCI is encoded at consecutive CCE indexes within a CORESET. The SS defines the \emph{starting} CCE index of all PDCCH \emph{candidate} locations carrying the DCI (UL or DL) for a specific UE for a given AL and at a given slot. Such indexes are generated via a hashing function that both the gNodeB and the UEs know and can compute. 
	The gNodeB selects only one candidate to encode the DCI addressed to a UE in UL or DL, that the UE is unaware of. Hence, the UE performs \emph{blind decoding}, i.e., it scans all available candidate locations for all possible ALs (since the UE also ignores its AL) until the DCI is successfully decoded. 
	To alleviate the complexity of blind decoding, the number $N_{AL}$ of candidates is limited and can be defined as a function of the AL. We remark that several CORESETs and SSs can be configured in each slot for each UE. 
	
	\emph{Related works.} PDCCH 3GPP specifications are detailed in \cite{TS38211} and disseminated by \cite{takeda2020understanding}. 
	The design of an efficient hashing function is investigated in \cite{braun20195g}. 
	The control channel scheduling has recently garnered attention for narrow-band Internet of Things (IoT) and RedCap devices. 
	There, the device energy consumption is a major bottleneck that can be mitigated by minimizing blind decoding \cite{yu2023energy} or by designing the SS \cite{manne2020scheduling} and the control period \cite{yu2024control}. 
	The problem of PDCCH candidate selection has already been formulated as a set packing problem, known to be equivalent to the independent set problem, in \cite{balamurali2010optimal} and a heuristic was proposed.
	
	In this paper we solve the PDCCH candidate selection via efficient combinatorial optimization techniques with performance guarantees.
	We demonstrate the superiority of our approach, based on combinatorial optimization theory, in terms of computational complexity and performance.

	\section{Problem formulation}
	
	We address the problem of selecting the PDCCH candidate on which the gNodeB transmits a DCI (for UL or DL) for each UE.
	Here we focus on a specific slot and we denote by $\mathcal U$ the set of UEs eligible for data transmission. We denote by $\mathcal C_u^{\mathrm{UL}},\mathcal C_u^{\mathrm{DL}}$ be the set of PDCCH candidates for UE $u\in \mathcal U$ across all active SSs and CORESETs, for UL and DL, respectively. For UE $u\in \mathcal U$, the gNodeB aims to select at most one candidate among $\mathcal C_u^{\mathrm{UL}}$ and at most one among $\mathcal C_u^{\mathrm{DL}}$. 
	Importantly, each CCE must only carry the DCI of a single UE. Equivalently, any two selected PDCCH candidates cannot overlap on any CCE.
	
	Our goal is to schedule as many DCIs as possible while avoiding overlaps. Moreover, we want to prioritize the allocation of UEs with higher scheduling priority. 
	We formalize these goals by assigning priority \emph{weights} $w_u^{\mathrm{UL}},w_u^{\mathrm{DL}}$ to UE $u\in \mathcal U$ in UL and DL, respectively, and by maximizing the sum of the weights of UEs with one selected candidate. 
	
	To define the optimization problem we introduce the auxiliary binary variable $x_c$ indicating whether ($x_c=1$) or not ($x_c=0$) a DCI is transmitted on candidate $c\in \mathcal C$. Note that $\mathcal C$ is the set of all possible candidates, i.e., $\mathcal C=\cup_{u\in\mathcal U} (\mathcal C_u^{\mathrm{UL}}\cup \mathcal C_u^{\mathrm{DL}})$. We then formulate our problem as:
	\begin{align}
		\max_{x \in\{0,1\}} & \, \sum_{u\in \mathcal U} w_u^{\mathrm{UL}} \sum_{c\in \mathcal C_u^{\mathrm{UL}}} x_c + \sum_{u\in \mathcal U} w_u^{\mathrm{DL}} \sum_{c\in \mathcal C_u^{\mathrm{DL}}} x_c \label{eq:obj}  \\
		\mathrm{s.t.} & \ x_c + x_{c'}\le 1 \qquad \forall \, \mathrm{overlapping} \ (c,c') \label{eq:no_conflicts} \\
		& \ \sum_{c\in \mathcal{C}^{i}_u} x_c \le 1, \quad \forall \, u\in\mathcal U, \ i\in\{\mathrm{UL},\mathrm{DL}\}.\label{eq:only_oneULDL}
	\end{align}
	
	Constraint \eqref{eq:no_conflicts} states that if two candidates overlap, then at most one of them can be selected, i.e., no pair of selected candidates can overlap. 
	According to \eqref{eq:only_oneULDL}, for each UE at most one candidates can be used in UL and DL, respectively.
	
	The optimization problem (\ref{eq:obj}-\ref{eq:only_oneULDL}) is a binary integer program. Specialized algorithms can be used to solve or approximate it by recognizing that it can be reformulated as a \emph{Maximum Weighted Independent Set} (MWIS) problem, as shown next. 
	
\begin{figure}
		\centering
		\includegraphics[width=1\linewidth]{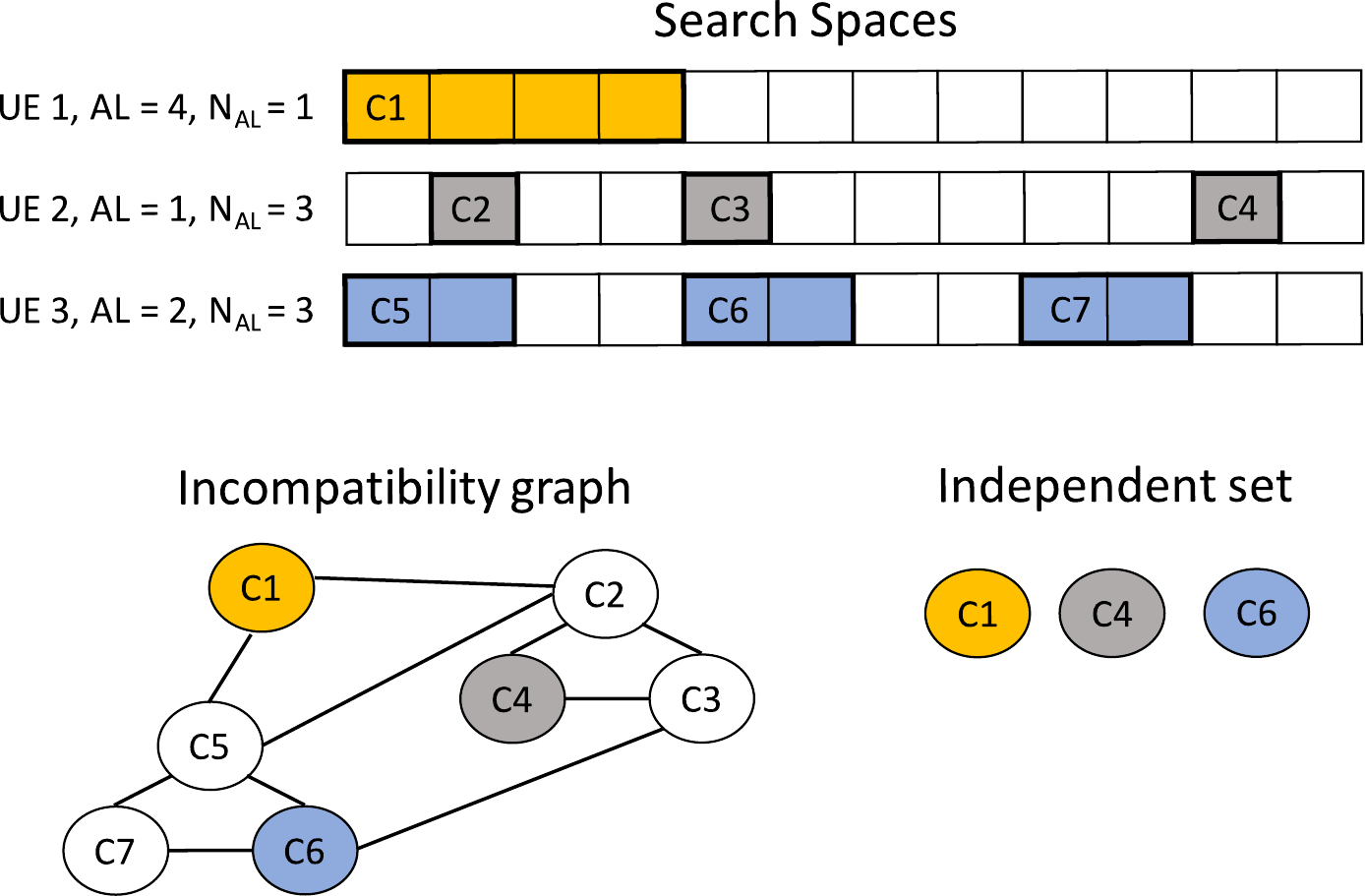}
		\caption{Incompatibility graph $G$ and an independent set for a given Search Spaces (SS), considering for simplicity exclusively DL or UL. There is no edge between any pair of vertices in the independent set. Equivalently, only non-overlapping candidates are chosen, and at most one per UE.}
		\label{fig:mwis_graph}
	\end{figure}

	The \textbf{MWIS formulation} of the PDCCH scheduling problem hinges on the definition of an auxiliary undirected weighted graph $G=(\mathcal C,E,W)$ that we call \emph{incompatibility graph}. Its vertices $\mathcal C$ are the set of PDCCH candidates. 
	There is an edge $(c,c')\in E$ if candidates $c,c'$ cannot be selected concurrently, i.e., if they overlap or are associated to the same UE in UL or DL, as shown in Figure \ref{fig:mwis_graph}. Formally, the set of edges is
	\begin{align}
		E=\Big\{ (c,c'): \, & \big[\exists \, (u,i): c\in \mathcal C^i_u \land c'\in \mathcal C^i_u \big] \lor \big[(c,c') \ \mathrm{overlap}\big] \Big \} \notag 
	\end{align}
	where $i\in \{\mathrm{UL},\mathrm{DL}\}$. Each node $c$ has weight $w_c\in W$ which, with some abuse of notation, is defined as $w_c=w_u^i$ if candidate $c$ belongs to UE $u$ and mode $i\in\{\mathrm{UL},\mathrm{DL}\}$. 
	
	We now rewrite the PDCCH scheduling problem (\ref{eq:obj}-\ref{eq:only_oneULDL}) as a MWIS on the incompatibility graph $G$. Here we aim to find the set of nodes with maximum weight and such that they form an \emph{independent set}, i.e., no pair of nodes shares an edge:
	\begin{align}
		\max_{x\in \{0,1\}} & \, \sum_{c\in \mathcal C} w_v x_v \label{eq:obj1}  \\
		\mathrm{s.t.} & \, x_c+x_{c'}\le 1, \qquad \forall \, (c,c')\in E. \label{eq:no_conflicts1} 
	\end{align}

	\section{Algorithms for PDCCH scheduling}

	The MWIS problem is known to be NP-hard, i.e., there exists no polynomial time algorithm that solves it exactly (unless P=NP). 
	Unfortunately, MWIS is also hard to approximate. In fact, there exists no polynomial time algorithm able to approximate it within a factor $\mathcal O(\mathrm{\#nodes^{\epsilon-1}})$ away from the optimal value, for $\epsilon>0$ \cite{williamson2011design}.
	However, several algorithms with performance guarantees have been proposed in the literature. 
	We select two methods from the MWIS literature that we deem suitable for PDCCH scheduling where the latency constraint is tight. 
	Then, we outline a procedure for computing the optimal solution without exhaustively enumerating all possibilities, although it remains impractical for our application. Finally, we describe the heuristic introduced in \cite{balamurali2010optimal}, requiring to compute the optimal solution on a small subset of users.

	\subsection{Greedy algorithm}
	
	The most computationally efficient algorithm studied in the literature for MWIS is a greedy one that iteratively selects the node (i.e., candidate) with the highest metric. Once a node is selected, all of its neighbors are subsequently removed from the graph, since selected nodes cannot share an edge.
	
	\begin{algorithm}
		Set $\mathcal C^{\mathrm{sel}}=\emptyset$ as the set of selected PDCCH candidates\;
		\While{$\mathcal C\ne \emptyset$}{
			Select candidate $c^*
			\in \mathcal C$ with highest metric $m_{c^*}$\;
			Add $c^*$ to $\mathcal C^{\mathrm{sel}}$\;
			Remove $c^*$ from incompatibility graph $G$ (and from $\mathcal C$) along with all of its neighbors\;
		}
		\caption{Greedy algorithm}
		\label{alg:greedy}
	\end{algorithm}
	
	Different definitions of metric $m$ lead to different variants of the greedy algorithm above. 
	A popular one, studied in \cite{sakai2003note}, is one where $m$ is the Weight-to-Degree Ratio (WDR):
	\begin{equation} \label{eq:metric1}
		m_c:=\frac{w_c}{\mathrm{deg}(c)+1}, \quad \forall\, c\in \mathcal C 
	\end{equation}
	where $\mathrm{deg}(c)$ is the degree, i.e., the number of neighbors, of node $c$. 
	This metric prioritizes the selection of candidates with higher weight and lower degree, thus blocking fewer other candidates. 
	Note that the WDR metric is recomputed after each iteration, since the graph is updated upon a node selection.
	
	\emph{Performance guarantees.} Under the WDR metric $m$ defined as in \eqref{eq:metric1}, the greedy algorithm produces a set of candidates with total weight of at least $\sum_{c \in \mathcal C} \frac{w_c}{\mathrm{deg}(c)+1}$, as shown in \cite{sakai2003note}.

	\subsection{Feige-Reichmann (FR) algorithm}
	
	The second algorithm for MWIS we present is named after its inventors Feige and Reichman \cite{feige2015recoverable}.
	It builds on the fact that computing the optimal solution for MWIS is easy for undirected \emph{forests}. (We recall that a forest is a collection of trees, and a tree is a graph with no cycle.)
	One can proceed iteratively from leaf to root nodes. 
	At each iteration the optimum values $V^Y(c), V^N(c)$ are available for lower subtrees rooted at $c$, whether $c$ is included ($V^Y(c)$) or not ($V^N(c)$) in the subtree optimal solution. 
	Then, such values are computed for the parent nodes, knowing that either the parent or the children, but not both, can be included in the optimal solution. 
	The complexity of the resulting procedure is \emph{linear} in the number of candidates.
	
	\begin{algorithm}
		\emph{Require}: Input graph $\underline{G}$ is an undirected forest\;
		Compute the value $V^Y(c)= w_\ell$ and $V^N(c) = 0$ for all leaf nodes $c$\;
		Until roots are reached, compute values for nodes $p$ as:
		\begin{align}
			V^Y(p) = & \, w_p + \sum_{c \in \mathrm{children}(p)} V^N(c) \\
			V^N(p) = & \, \sum_{c \in \mathrm{children}(p)} \max\{V^N(c), V^Y(c) \}
		\end{align}
		\textbf{return} optimal value $\sum_{r \in \mathrm{roots}} \max \left\{ V^N\!(r), \, V^Y\!(r) \right\}$
		\caption{Optimal MWIS for forests}
		\label{alg:mwis-forest}
	\end{algorithm}
	
	
	Note that the optimal subtree \emph{solution} can be computed iteratively using the same approach as for calculating the value.
	
	The Feige-Reichmann (FR) algorithm first extracts a forest as a subgraph of the incompatibility graph $G$ and then uses Algorithm \ref{alg:mwis-forest} as a subroutine. 
	The forest is produced by iterating over all nodes in a certain order and adding a node only if at most one neighbor was encountered previously. The resulting subgraph cannot contains cycles: in fact, if a cycle were to exist, its last added node would have at least two neighbors among the previous nodes, leading to a contradiction.
	
	\emph{Performance guarantees.}  The FR procedure always produces a feasible solution; yet, its quality depends on the node sorting criterion used in line 2, Algorithm \ref{alg:feig-reich}.  
	If the candidates are sorted via a uniformly random permutation, then the \emph{expected} sum of weights of the produced solution is at least $2\sum_{c \in \mathcal C^*} \frac{w_c}{\mathrm{deg}(c)+1}$, where $\mathcal C^*$ is the optimal solution \cite{feige2015recoverable}. Yet, the \emph{variance} in the solution quality can be an issue in practical PDCCH applications. 
	
	A safer approach would be to sort candidates by first including those produced by the greedy algorithm. The resulting solution shall be at least as good as the greedy one: indeed, the forest shall include the greedy solution since it forms an independent set, i.e., no edges are present among those nodes.
	
	\begin{algorithm}
		Initialize $\underline{G}$ as the empty graph\;
		Sort nodes $\mathcal C$ according to a certain criterion\;
		\For{\textnormal{candidate $c$ in the sorting order}}{
			If at most one neighbor $c'$ of $c$ was already seen, then add $c$ to $\underline{G}$ along with edge $(c,c')$, if any\;
		}
		Solve MWIS optimally on $\underline{G}$ via Algorithm \ref{alg:mwis-forest}
		\caption{Feige-Reichmann (FR) algorithm \cite{feige2015recoverable}}
		\label{alg:feig-reich}
	\end{algorithm}

	\subsection{Exact solution via recursion}
	
	To compute the optimal MWIS solution, a naive procedure requiring $\mathcal O(2^{|V|})$ steps would enumerate all subsets of nodes and check if they form an independent set.
	A better procedure reported below exploits a similar recursive property used by the FR algorithm and runs in $\mathcal O(1.38^{|V|})$ steps \cite{cormen2009introduction}.

	\begin{algorithm}
		\textbf{If} $G$ is empty \textbf{then return} 0 \\
		\Else{
			Choose a node $p$ \;
			Compute $V^N(p)=\text{Optimal-Recursion}(G \setminus \{p\})$\;
			Compute $V^Y(p)=w_p + \text{Optimal-Recursion}(G \setminus \{p\} \setminus \mathrm{neighbors}(p))$\;
			\textbf{return} $\max\left\{V^Y(p),V^N(p)\right\}$
		} 
		\caption{Optimal-Recursion($G$)}
		\label{alg:Recursion}
	\end{algorithm}
	
	\subsection{Optimal-then-Greedy (OtG) heuristic \cite{balamurali2010optimal}} \label{sec:soa}
	
	The exponential complexity of the optimal solution is prohibitive for our applications. For this reason, \cite{balamurali2010optimal} proposes a heuristic that first sorts UEs in descending order of metric $w/AL$. Then, the optimal solution is computed \emph{only} on the top $M$ UEs. Finally, the remaining candidates are selected via a greedy procedure using the same metric. 
	The complexity of the overall procedure is exponential in $M$. We found that only values $M\le 4$ are feasible in our latency-constrained scenario.

	\begin{table}[h]\footnotesize 
		\caption{Performance of PDCCH candidate selection methods}
		\label{tab:results}
		\begin{center}
			\begin{tabular}{|  L{1.47cm} | C{.8cm} | C{.76cm} | C{.76cm} | C{.41cm} | C{.41cm} | C{.41cm} | C{.41cm}|}
				\hline
				\multicolumn{4}{|c}{} & \multicolumn{4}{|c|}{\bf{inter-tx $\#$slots}} \\ \hline 
				\bf{Method} & \!\!\!\!\bf{$\#$UE/slot} & \!\!\!\bf{geomean thpt} & \!\!\bf{runtime} & \!\!\!\bf{AL=1} & \!\!\!\bf{AL=2} & \!\!\!\bf{AL=4} & \!\!\!\bf{AL=8} \\ 
				\hline
				\textbf{W-Greedy}            & \textcolor{red}{\bf 11.79} & 0.94 & \textcolor{ForestGreen}{\bf 0.43} & 1.36 & 2.34 & 4.18 & 6.52 \\
				\hline 
				\textbf{OtG}   & 13.15 & \textcolor{red}{\bf 0.84} & \textcolor{red}{\bf 13.52} & 1.03 & 1.97 & \textcolor{red}{\bf 5.91} & \!\!\!\textcolor{red}{\bf 12.19} \\
				\hline 
				\textbf{\mbox{WDR-Greedy}}  & \textbf{13.55}  & \textbf{1.00} & \textbf{1.00} & 1.12 & 1.94 & 4.03 & 8.07 \\
				\hline 
				\textbf{FR}  & \textcolor{ForestGreen}{\bf 13.74}  & \textcolor{ForestGreen}{\bf 1.01} & 3.32 & 1.10 & 1.89 & 4.08 & 8.27 \\
				\hline
				
			\end{tabular}
			
		\end{center}
		
	\end{table}

	\section{Numerical evaluations}
	
	We evaluated the performance of different options for PDCCH candidate selection under the following settings.
	The number of candidates per AL, $N_{AL}$, is set to $N_{AL=1}=5$, $N_{AL=2}=4$, $N_{AL=4}=3$, $N_{AL=8}=2$. 
	The same CORESET composed of 32 CCEs is used for all UEs.
	The UE-specific SS are generated according to the hashing function defined by 3GPP in \cite{TS38211}.
	Weights $w$ are defined as the classic Proportional Fairness (PF) metric, which is the ratio of the currently achievable rate to the average achieved throughput in the past. 
	For simplicity we assume that each UE has only one DCI, in DL.
	We ran our experiments on 100 independent scenarios. In each scenario, 30 UEs have a different AL randomly drawn from $\{1,2,4,8\}$ according to a uniform distribution. The AL remains unchanged across all 300 slots.
	Traffic requests are generated via a Poisson distribution, while spectral efficiency is inversely proportional to the AL. 
	In each slot, the PDSCH resources are distributed in round robin fashion across the UEs whose DCI has been assigned to a PDCCH candidate.
	
	\emph{Tested algorithms.} 
	We compared four different algorithms for PDCCH candidate selection. 
	The first one, called \emph{W-Greedy}, corresponds to the greedy Algorithm \ref{alg:greedy} with a simple \emph{w}eight node metric $m:=w$. Although easy to implement as it does not rely on the incompatibility graph, W-Greedy is myopic since it fails to consider how a candidate selection may subsequently block other UEs. 
	The second option is Optimal-then-Greedy (\emph{OtG}), proposed in \cite{balamurali2010optimal} and described in Section \ref{sec:soa}, where $M=4$. Higher values of $M$ proved computationally prohibitive.
	The third method, \emph{WDR-Greedy}, is greedy Algorithm \ref{alg:greedy} under WDR metric $m$ defined in \eqref{eq:metric1} and studied in \cite{sakai2003note}.
	The fourth solution is Feige-Reichmann (\emph{FR}) Algorithm \ref{alg:feig-reich} that uses WDR-Greedy as a sub-routine. In fact, the candidates produced by WDR-Greedy are placed on top of the list; then, the remaining candidates are sorted according to the WDR metric. In this setting, FR will always performs at least as well as WDR-Greedy, but with higher complexity.

	\emph{Results.} 
	We present the outcome of our numerical evaluations in Table \ref{tab:results}, including the number of scheduled UEs per slot, the geometric mean of DL throughput across all UEs and the running time (both normalized with respect to WDR-Greedy) as well as the number of slots between consecutive transmissions for UEs with different AL values. These values are averaged across the 100 simulated scenarios.
	
	As anticipated, the FR algorithm emerges as the top-performing method, capable of scheduling the highest number of UEs per slot and achieving the largest geometric mean throughput. WDR-Greedy algorithm closely follows on both metrics, trailing FR by only 1\%. 
	In contrast, W-Greedy is myopic and leads to a low number of scheduled UEs and, consequently, low throughput fairness.
	The OtG method presented in \cite{balamurali2010optimal} schedules a reasonable number of UEs per slot but exhibits low performance in terms of throughput fairness.
	Specifically, UEs with high AL are less frequently scheduled, as evidenced by the large inter-transmission number of slots for AL$=4,8$.
	We attribute this discrepancy to the node ranking metric, which is calculated as the ratio of weight to AL and excessively penalizes candidates with high AL. 
	Rectifying this issue would require a substantial increase in $M$, which becomes impractical due to complexity constraints.
	
	When also examining execution times reported in Table \ref{tab:results}, WDR-Greedy stands out as offering the most favorable performance-to-complexity trade-off. In fact, WDR-Greedy is three times faster than the top-performing FR with only a 1\% performance degradation, as mentioned earlier. W-Greedy is the lightest option as it does not leverage the incompatibility graph, while OtG is the most complex method as it involves calling the Optimal-Recursion Algorithm \ref{alg:Recursion}.
	

	\section{Conclusions}
	
	To ease the burden of blind decoding at the UE side, the gNodeB encodes the DCIs over a limited number of candidate frequency locations. The resulting PDCCH candidate selection problem faced by the gNodeB is NP-hard and even impossible to approximate.
	We addressed the PDCCH candidate selection problem via techniques rooted in combinatorial optimization theory. One of them, called WDR-Greedy, stands out as offering an excellent performance-vs-complexity trade-off, making it suitable for implementation in 5G NR schedulers.
	


\begin{thebibliography}{10}
		\providecommand{\url}[1]{#1}
		\csname url@samestyle\endcsname
		\providecommand{\newblock}{\relax}
		\providecommand{\bibinfo}[2]{#2}
		\providecommand{\BIBentrySTDinterwordspacing}{\spaceskip=0pt\relax}
		\providecommand{\BIBentryALTinterwordstretchfactor}{4}
		\providecommand{\BIBentryALTinterwordspacing}{\spaceskip=\fontdimen2\font plus
			\BIBentryALTinterwordstretchfactor\fontdimen3\font minus
			\fontdimen4\font\relax}
		\providecommand{\BIBforeignlanguage}[2]{{%
				\expandafter\ifx\csname l@#1\endcsname\relax
				\typeout{** WARNING: IEEEtran.bst: No hyphenation pattern has been}%
				\typeout{** loaded for the language `#1'. Using the pattern for}%
				\typeout{** the default language instead.}%
				\else
				\language=\csname l@#1\endcsname
				\fi
				#2}}
		\providecommand{\BIBdecl}{\relax}
		\BIBdecl
		
		\bibitem{TS38211}
		3GPP, ``{NR Physical channels and modulation},'' 3GPP, Tech. Rep. TS38.211,
		2018.
		
		\bibitem{takeda2020understanding}
		K.~Takeda, H.~Xu, T.~Kim, K.~Schober, and X.~Lin, ``{Understanding the heart of
			the 5G air interface: An overview of physical downlink control channel for 5G
			new radio},'' \emph{IEEE Communications Standards Magazine}, vol.~4, no.~3,
		pp. 22--29, 2020.
		
		\bibitem{braun20195g}
		V.~Braun, K.~Schober, and E.~Tiirola, ``{5G NR physical downlink control
			channel: Design, performance and enhancements},'' in \emph{2019 IEEE Wireless
			Communications and Networking Conference (WCNC)}.\hskip 1em plus 0.5em minus
		0.4em\relax IEEE, 2019, pp. 1--6.
		
		\bibitem{yu2023energy}
		Y.-J. Yu and C.-L. Wu, ``{Energy-Efficient Scheduling for Search-Space Periods
			in NB-IoT Networks},'' \emph{IEEE Systems Journal}, 2023.
		
		\bibitem{manne2020scheduling}
		P.~R. Manne, S.~Ganji, A.~Kumar, and K.~Kuchi, ``{Scheduling and decoding of
			downlink control channel in 3GPP narrowband-IoT},'' \emph{IEEE Access},
		vol.~8, pp. 175\,612--175\,624, 2020.
		
		\bibitem{yu2024control}
		Y.-J. Yu, Y.-C. Wang, and C.-H. Fan, ``Control period adaptation and resource
		allocation for joint uplink and downlink in nb-iot networks,'' \emph{IEEE
			Internet of Things Journal}, 2024.
		
		\bibitem{balamurali2010optimal}
		Balamurali, ``{Optimal downlink control channel resource allocation for LTE
			systems},'' in \emph{2010 International Conference on Signal Processing and
			Communications (SPCOM)}.\hskip 1em plus 0.5em minus 0.4em\relax IEEE, 2010,
		pp. 1--5.
		
		\bibitem{williamson2011design}
		D.~P. Williamson and D.~B. Shmoys, \emph{The design of approximation
			algorithms}.\hskip 1em plus 0.5em minus 0.4em\relax Cambridge university
		press, 2011.
		
		\bibitem{sakai2003note}
		S.~Sakai, M.~Togasaki, and K.~Yamazaki, ``A note on greedy algorithms for the
		maximum weighted independent set problem,'' \emph{Discrete applied
			mathematics}, vol. 126, no. 2-3, pp. 313--322, 2003.
		
		\bibitem{feige2015recoverable}
		U.~Feige and D.~Reichman, ``Recoverable values for independent sets,''
		\emph{Random Structures \& Algorithms}, vol.~46, no.~1, pp. 142--159, 2015.
		
		\bibitem{cormen2009introduction}
		T.~H. Cormen, C.~E. Leiserson, R.~L. Rivest, and C.~Stein, \emph{{Introduction
				to Algorithms}}.\hskip 1em plus 0.5em minus 0.4em\relax MIT press, 2009.
		
	\end{thebibliography}

\end{document}